\documentclass[aps, prb, reprint, longbibliography]{revtex4-2}
\usepackage{graphicx}% Include figure files
\usepackage{dcolumn}% Align table columns on decimal point
\usepackage{color}
\usepackage{braket}
\usepackage{amsmath,amssymb,amsfonts,bm}
\usepackage[normalem]{ulem}

\def\mk{\mathbf{k}}
\def\mq{\mathbf{q}}

\begin{document}

\title{Phonon-assisted photoluminescence of bilayer MoS$_2$ from first principles}
\author{Gyanu P. Kafle}
\author{Zhen-Fei Liu}
\email{zfliu@wayne.edu}
\affiliation{Department of Chemistry, Wayne State University, Detroit, Michigan 48202 USA}

\date{\today}

\begin{abstract}
In indirect band gap materials, phonon-assisted processes are key mechanisms for photoluminescence (PL). Using a first-principles many-body approach, we systematically investigate the phonon-assisted PL in bilayer MoS$_2$ and its dependence on temperature and external tensile strain. The effects of phonons are accounted for using a supercell approach: we identify the phonon momenta that are important to PL, construct supercells that are commensurate with these phonons, and examine the changes in the optical absorption after explicit displacements of atoms along each phonon mode. The PL intensity is then obtained via the van Roosbroeck-Shockley relationship from the optical absorption spectra. This approach enables us to investigate phonon-absorption and phonon-emission processes separately and how each process depends on temperature. Our results reveal that optical phonons associated with out-of-plane vibrations of S atoms and in-plane vibrations of Mo atoms contribute most to the indirect PL for unstrained bilayer MoS$_2$. Additionally, we also discuss how the PL spectra and the phonon contributions evolve with strain. In particular, we show that at high strain, additional phonon channels become available due to the modulation of the electronic band structure. 
\end{abstract}

\maketitle

\section{Introduction}
% TMDs optical properties 
Transition-metal dichalcogenides possess rich electronic and optical properties tunable by thickness and strain. Using MoS$_2$ as an example, the monolayer is a direct-gap semiconductor \cite{Mak2010,Splendiani2010,Qiu2016, Hippalgaonkar2017}, making it a highly efficient light emitter with pronounced photoluminescence (PL) through the direct recombination of excitons \cite{Mak2010, Splendiani2010}. In contrast, bilayer and multilayer MoS$_2$ are indirect-gap semiconductors \cite{Cheiwchanchamnangij2012, Hippalgaonkar2017}, resulting in significantly weaker PL intensities \cite{Conley2013} due to the mismatch in momentum between the electron and hole states. In these materials, PL spectra often display multiple peaks that are assigned to both direct and indirect transitions \cite{Mak2010,Conley2013}. Furthermore, it was shown that the PL intensity in bilayer MoS$_2$ can be suppressed by a relatively small uniaxial tensile strain \cite{Conley2013,Zhu2013}. In addition, Refs. \cite{Du2018} and \cite{Pei2015} showed that the PL efficiency and spectral features can be modulated by temperature, where the indirect emission exhibits reduced intensity and a redshift as the temperature decreases. These experimental observations highlight the need for a comprehensive theoretical framework to understand how structural and environmental factors affect excitonic and PL behaviors, which are pivotal in advancing their applications in optoelectronics.

% theory of PL from first principles 
In indirect-gap materials, phonons play a crucial role in PL by providing the momentum required for radiative recombination. Under certain conditions, phonon-assisted optical transitions give rise to significant PL \cite{Mak2010, Splendiani2010}, with a well-known example being hexagonal boron nitride (hBN) \cite{Cassabois2016, Paleari2019}. A unified theory of PL should capture both direct (phonon-less) and indirect (phonon-assisted) electronic transitions \cite{Tiwari2024}. If one focuses on the indirect transitions, a proper description of the exciton-phonon coupling is required for materials with significant exciton effects. To this end, the computational approaches largely fall into two categories. In the finite-displacement method, the transition matrix elements are evaluated under perturbation with explicit atomic distortions along phonon modes \cite{Paleari2019, Cannuccia2019}. This approach requires the construction of supercells commensurate with relevant phonon wave vectors. It is straightforward to implement and provides a transparent and intuitive way to describe the effects of phonons. Alternatively, the exciton-phonon coupling can be calculated by combining density-functional perturbation theory (DFPT) for electron-phonon interactions with the Bethe-Salpeter equation (BSE) for excitons \cite{Chen2020, Cudazzo2020, Antonius2022, Paleari2022}, which is a key ingredient in computing PL \cite{lechifflart2023, Zanfrognini2023, Marini2024}. This approach avoids the construction of large supercells, but requires significant implementation. 

%In this work
In this work, we largely follow the theoretical framework proposed in Ref. \cite{Paleari2019} to study the PL in bilayer MoS$_2$ using first principles, as well as its temperature and strain dependence. The electronic and optical properties are described at the level of many-body perturbation theory using the $GW$-BSE approach ($G$: Green’s function; $W$: screened Coulomb interaction) \cite{HL86, Rohlfing2000}, which has been shown to be essential for MoS$_2$ \cite{Cheiwchanchamnangij2012, Qiu2016, Hippalgaonkar2017}. To incorporate the exciton-phonon coupling, we explicitly distort the atomic coordinates along phonon eigenvectors in a finite-displacement approach, in supercells that are commensurate with the relevant phonon modes. While Ref. \cite{Paleari2019} only focused on PL processes with phonon emission due to the low temperature considered, here we examine PL mechanisms involving both phonon absorption and phonon emission, and systematically analyze how temperature modulates each pathway. Moreover, we decompose the PL spectra to contributions from each phonon mode and elucidate how the PL evolves with strain in this indirect-gap material. This comprehensive approach provides insight into PL in bilayer MoS$_2$, highlighting the critical role of phonon-assisted processes and the effects of strain and temperature. 

This paper is structured as follows. Sec. \ref{sec:theory} presents the basic theoretical framework of phonon-assisted PL. Sec. \ref{sec:supercell} discusses the specific techniques we use to incorporate the effects of phonons in the finite-displacement approach. Sec. \ref{sec:comppara} lists the computational parameters in detail. Sec. \ref{sec:mos2} presents the PL results of unstrained MoS$_2$ bilayer at room temperature. Sec. \ref{sec:temp-dep} examines the temperature dependence and Sec. \ref{sec:strain} examines the strain dependence. We conclude this work in Sec. \ref{sec:conclude}.

\section{Methods}

\subsection{Theory of phonon-assisted PL}
\label{sec:theory}
In this section, we outline the major steps in computing PL from first-principles $GW$-BSE, largely following Ref. \cite{Paleari2019}. The basic idea is to first express the rate of spontaneous emission, $R_{\rm sp}(\omega)$, in terms of the absorption coefficient, $\alpha(\omega)$, using the van Roosbroeck-Shockley relation \cite{Roosbroek1954,Bebb1972}. Then, $\alpha(\omega)$ can be expressed using the imaginary part of the macroscopic dielectric function $\varepsilon_2(\omega)$, which is related to the transition matrix elements that can be computed in the BSE formalism. The effects of phonons are taken into account by (i) differentiating the optical absorption and emission frequencies in the formalism and (ii) considering the Taylor expansion of $\varepsilon_2(\omega)$ in phonon modes. We use atomic units throughout this work, i.e., $\hbar=1/(4\pi\varepsilon_0)=e=m_e=1$. 

The van Roosbroeck-Shockley relation \cite{Roosbroek1954,Bebb1972} is based on the detailed balance between light absorption and emission in thermal equilibrium. Considering phonon-assisted processes, it reads
\begin{equation}
\begin{split}
R_{\rm sp}^{\pm}(\omega_e) = \frac{n_r(\omega_e)\omega_e}{n_r(\omega_a)\omega_a} & V_g(\omega_a)G(\omega_a)\alpha^\pm(\omega_a) \\
& \times\exp{\left(-\frac{\bar{\omega}-\Delta\mu}{k_BT}\right)}.
\end{split}
\label{eq:rs}
\end{equation}
The total PL rate is $R_{\rm sp}(\omega_e) = R_{\rm sp}^+(\omega_e) + R_{\rm sp}^-(\omega_e)$. Here, $\omega_e$ ($\omega_a$) is the frequency for light emission (absorption). $R_{\rm sp}$ is the rate of spontaneous light emission and $\alpha$ is the light absorption coefficient. In both quantities, the $+$ ($-$) sign indicates processes involving phonon emission (absorption). We have $\omega_a=\omega_e\pm2\Omega$ \cite{Bebb1972}, with $\Omega$ the phonon frequency, and have defined $\bar{\omega}=(\omega_e+\omega_a)/2$, the frequency of the phonon-less electronic transition. $n_r$ is the refractive index, $V_g(\omega_a)$ is the group velocity of the photons, and $G(\omega_a)$ is the photon density of states given by
\begin{equation}
G(\omega_a)=\frac{n_r^2(\omega_a)\omega_a^2}{(\pi c)^2V_g(\omega_a)},
\label{eq:G}
\end{equation}
where $c$ is the speed of light in vacuum. In this work, we consider $n_r(\omega_a)\approx 1$ and $n_r(\omega_e)\approx 1$, i.e., the refractive index is weakly dependent on energy in the energy range of interest. In the last factor in Eq. \eqref{eq:rs}, $\Delta\mu=\mu_{\rm e}-\mu_{\rm h}$, the difference in the chemical potentials of the electrons and holes. $k_B$ is the Boltzmann constant and $T$ is the temperature. In phonon-less processes, $\omega_e=\omega_a$ and Eq. \eqref{eq:rs} is reduced to the standard van Roosbroeck-Shockley form \cite{Roosbroek1954,Bebb1972}.

The absorption coefficient, $\alpha(\omega_a)$, can be expressed in terms of the imaginary part of the macroscopic dielectric function, $\varepsilon_2(\omega_a)$ \cite{Yu-cardona}:
\begin{equation}
\alpha(\omega_a)=\frac{\omega_a}{cn_r(\omega_a)}\varepsilon_2(\omega_a).
\label{eq:alpha}
\end{equation}
Again, we consider the $n_r(\omega_a)\approx 1$ limit in Eq. \eqref{eq:alpha} in this work. The modification of this equation due to phonons will be discussed below.

For phonon-assisted processes in materials with indirect band gaps, we consider the Taylor expansion of the macroscopic dielectric function, $\varepsilon_2(\omega_a)$, up to the second order \cite{Zacharias2016}:
\begin{equation}
\varepsilon_2(\omega_a)= \varepsilon_2^{(0)}(\omega_a) + \varepsilon_2^{(2)}(\omega_a).
\label{eq:taylor}
\end{equation}
The first-order term $\varepsilon_2^{(1)}$ in Eq. \eqref{eq:taylor} vanishes when we take the thermal average. Substituting this equation to Eq. \eqref{eq:alpha} gives rise to two terms, one depending on $\varepsilon_2^{(0)}(\omega_a)$ and another depending on $\varepsilon_2^{(2)}(\omega_a)$. The former corresponds to PL without phonon contributions, and the latter corresponds to the phonon-assisted PL, the focus of this work. In the rest of the discussion, we only consider the $R_{\rm sp}^\pm(\omega_e)$ expressions that depend on $\varepsilon_2^{(2)}(\omega_a)$. 

The $\varepsilon_2^{(2)}(\omega_a)$ in Eq. \eqref{eq:taylor} can be expressed by \cite{Zacharias2016}
\begin{equation}
\varepsilon_{2}^{(2)}(\omega_a) = \frac{1}{2} \sum_{\lambda\mathbf{q}} \left. \frac{\partial^2 \varepsilon_{2}^{(0)}(\omega_a)}{\partial R_{\lambda\mathbf{q}}^2} \right|_{\rm eq} l_{\lambda\mathbf{q}}^2\left(2n_B+1\right).
\label{eq:eps-2nd-order}
\end{equation}
Here, ``eq'' denotes that the second derivative is evaluated at the equilibrium structure ($R_{\lambda\mq}=0$). $\lambda$ is a phonon mode and $\mq$ is a phonon wave vector in the first Brillouin zone. $R_{\lambda\mathbf{q}}$ is the distortion of atomic coordinates along the phonon mode $\lambda$ at wave vector $\mathbf{q}$. The factors after the second derivative come from the thermal average \cite{Patrick2014}, where $l_{\lambda\mathbf{q}}^2= 1/(2 M_{\lambda\mathbf{q}}\Omega_{\lambda\mathbf{q}})$, with $M_{\lambda\mathbf{q}}$ being the weighted unit-cell mass and $\Omega_{\lambda\mathbf{q}}$ the phonon frequency. $n_B$ is the Bose-Einstein occupation for phonons:
\begin{equation}
n_B(\Omega_{\lambda\mq}, T) = \frac{1}{\exp{\left(\frac{\Omega_{\lambda\mathbf{q}}} {k_\text{B} T}\right)} - 1}. 
\label{eq:bose-occu}
\end{equation}

In the exciton picture, $\varepsilon_2^{(0)}(\omega_a)$ can be computed using the BSE \cite{Rohlfing2000}, involving transition matrix elements between the ground state and excited states:
\begin{equation}
\varepsilon_2^{(0)}(\omega_a)= 4\pi^2\sum_S \left|T^S\right|^2 \delta( \omega_a - E^S),
\label{eq:bse}
\end{equation}
where $T^S=\braket{0|\mathbf{e}\cdot\mathbf{v}|S}/\omega_a$, the velocity matrix element between ground state $\ket{0}$ and an excited state $\ket{S}$ along the direction of light polarization $\mathbf{e}$. $E^S$ is the excitation energy. For simplicity, we have omitted the $\mathbf{k}$-point dependence in Eq. \eqref{eq:bse}. In an independent particle picture \cite{Yu-cardona}, the sum over excited states $\sum_{S}$ is reduced to a sum over $v\to c$ transitions ($v$: valence band; $c$: conduction band), $\sum_{vc}$. $T$ then becomes $\braket{c|\mathbf{e}\cdot\mathbf{v}|v}/\omega_a$, and $E^S$ becomes $\epsilon_c-\epsilon_v$.

Taking the second derivative of Eq. \eqref{eq:bse} and using the fact that $T^S=0$ for the indirect transition of the equilibrium structure, we can rewrite Eq. \eqref{eq:eps-2nd-order} as
\begin{equation}
\begin{split}
\varepsilon_{2}^{(2)}(\omega_a) = 2\pi^2\sum_{\lambda\mathbf{q}}\sum_S & \left.\frac{\partial^2|T^S|^2}{\partial R_{\lambda\mathbf{q}}^2}\right|_{\rm eq} l_{\lambda\mathbf{q}}^2 \\
& \times \delta (\omega_a - E^S)\left(2n_B+1\right).
\end{split}
\label{eq:eps-2nd-order2}
\end{equation}

For phonon-assisted transitions, we replace the factors in the second line of Eq. \eqref{eq:eps-2nd-order2} by the following:
\begin{equation}
\begin{split}
\delta (\omega_a - E^S)(2n_B + 1) & \Rightarrow \delta (\omega_a - [E^S-\Omega_{\lambda\mathbf{q}}]) n_B \\
&+ \delta (\omega_a - [E^S+\Omega_{\lambda\mathbf{q}}])(n_B + 1). \\
\end{split}
\label{eq:ratio}
\end{equation} 
Here, the first term represents phonon absorption in light-absorption processes: the excitation energy is red-shifted due to the participation of a phonon, and the phonon absorption corresponds to the annihilation of a phonon and occurs with a probability proportional to $n_B$. Similarly, the second term represents phonon emission in light-absorption processes: the excitation energy is blue-shifted, and the phonon emission corresponds to the creation of a phonon and occurs with a probability proportional to $n_B + 1$. 

In phonon-assisted PL, we substitute the first term of Eq. \eqref{eq:ratio} into Eq. \eqref{eq:eps-2nd-order2}, then substitute it into Eq. \eqref{eq:alpha} to obtain $\alpha^-(\omega_a)$. Together with Eq. \eqref{eq:G}, the result is substituted into Eq. \eqref{eq:rs} to obtain $R_{\rm sp}^-$, which represents PL rate with phonon absorption:
\begin{equation}
\begin{split}
R_{\rm sp}^-(\omega) = & \sum_{\lambda\mathbf{q}} \frac{2\omega(\omega - 2\Omega_{\lambda\mathbf{q}})^2}{c^3} l_{\lambda\mathbf{q}}^2 \sum_S  \left. \frac{\partial^2 |T^S|^2}{\partial R_{\lambda\mathbf{q}}^2} \right|_{\rm eq} \\
& \times \delta(\omega - E^S - \Omega_{\lambda\mathbf{q}})n_B N_{\rm xct},
\end{split}
\label{eq:rsp-}
\end{equation}
where we have used $\omega_a = \omega_e-2\Omega_{\lambda\mq}$ and then rewritten $\omega_e$ simply as $\omega$. 

Similarly, we substitute the second term of Eq. \eqref{eq:ratio} into Eqs. \eqref{eq:eps-2nd-order2}, \eqref{eq:alpha}, and \eqref{eq:rs} to obtain $R_{\rm sp}^+$, which represents PL rate with phonon emission: 
\begin{equation}
\begin{split}
R_{\rm sp}^+(\omega) = & \sum_{\lambda\mathbf{q}} \frac{2\omega(\omega + 2\Omega_{\lambda\mathbf{q}})^2}{c^3}  l_{\lambda\mathbf{q}}^2 \sum_S  \left. \frac{\partial^2 |T^S|^2}{\partial R_{\lambda\mathbf{q}}^2} \right|_{\rm eq}\\
& \times \delta(\omega - E^S + \Omega_{\lambda\mathbf{q}})(n_B+1) N_{\rm xct},
\end{split}
\label{eq:rsp+}
\end{equation}
where we have used $\omega_a = \omega_e+2\Omega_{\lambda\mq}$ and then rewritten $\omega_e$ simply as $\omega$. In both Eq. \eqref{eq:rsp-} and Eq. \eqref{eq:rsp+}, we have also defined $N_{\rm xct}$ and followed Ref. \cite{Paleari2019} to make the approximation
\begin{equation}
\begin{split}
\exp{\left(-\frac{\bar{\omega}-\Delta\mu}{k_BT}\right)} & := N_{\rm xct}(E^S,T) \\
& \approx \exp\left(-\frac{E^S - E^{S1}}{k_\text{B} T}\right).
\end{split}
\label{eq:xct-occu}
\end{equation}

The rationale behind Eq. \eqref{eq:xct-occu} is the following: the populations of the electrons and holes are approximated by an equilibrium Boltzmann distribution of excitons, $N_{\rm xct}$, which depends parametrically on the excitation energy $E^S$. $\Delta\mu$ can be understood as a chemical potential-like quantity that is chosen to be the energy of the lowest bound exciton $E^{S1}$ \cite{Paleari2019}. We also note in passing that in principle, the nominal exciton temperature appearing in the second line of Eq. \eqref{eq:xct-occu} might be different from the lattice temperature appearing in the first line of Eq. \eqref{eq:xct-occu} \cite{Cassabois2016, Paleari2019, lechifflart2023}. However, the relationship between the two temperatures is often material-dependent.

%This choice is motivated by the fact that, below the indirect gap, optical transitions occur not into a continuum of states but into a discrete set of excitonic states with well-defined energies. Consequently, the exciton occupation function $N_{\mathrm{xct}}$ is nonzero only when the photon-phonon combined energy, $E^{\pm} = \omega \pm \Omega_{\lambda\mathbf{q}}$, matches an exciton energy $E^S$. In other words, $N_{\mathrm{xct}}$ contributes only when energy conservation is satisfied for a transition into a specific excitonic state via phonon absorption and emission. In this work, $N_{\mathrm{xct}}$ can be approximated by the Bose-like occupation of excitons,

Finally, the second derivatives of the transition matrix elements in Eqs. \eqref{eq:rsp-} and \eqref{eq:rsp+} are approximated using the finite-difference method:

\begin{equation}
\begin{aligned}
& \left.\frac{\partial^2\left|T^S\right|^2}{\partial R_{\lambda\mathbf{q}}^2}\right|_{\rm eq} \\
& = \frac{\left|T^S (R_{\lambda\mq})\right|^2 + \left|T^S (-R_{\lambda\mq})\right|^2- 2\left|T^S (R_{\lambda \mq}=0)\right|^2}{A^2} \\
&\approx 2\frac{\left|T^S (R_{\lambda\mq})\right|^2}{A^2},
\end{aligned}
\label{eq:finite-diff}
\end{equation}
where we have neglected $|T^S (R_{\lambda \mq}=0)|^2$ \cite{Paleari2019} in the third line. This is because for the indirect-gap undistorted structure $(R_{\lambda \mq}=0)$, $T^S$ is negligible. In fact, in our numerical calculations, it is several orders of magnitude smaller than $T^S(R_{\lambda\mq})$. This approximation also leads to $\partial |T^S|^2/\partial R_{\lambda\mq}=0$ for the equilibrium structure, such that $|T^S(R_{\lambda\mq})|^2 = |T^S(-R_{\lambda\mq})|^2$. The $A$ in the denominator is the amplitude of the distortion [see Eq. \eqref{eq:distort} below].

In the remaining parts of the paper, we compute $R_{\rm sp}(\omega) = R_{\rm sp}^+(\omega) + R_{\rm sp}^-(\omega)$ using Eqs. \eqref{eq:rsp-} and \eqref{eq:rsp+}, and analyze how $R_{\rm sp}(\omega)$ and each of its two components change with temperature.

%We employed the same temperature ($T$) in Eqs. \eqref{eq:bose-occu} and  \eqref{eq:xct-occu2} when presenting our results, rather than using the lattice temperature in Eq. \eqref{eq:bose-occu} and excitonic temperature in Eq. \eqref{eq:xct-occu2} as adopted in Refs.  for hBN. This choice is motivated by the lack of experimental data specific to the variation of excitonic temperature in MoS$_2$ bilayer. While studies such as Ref.~\cite{Cassabois2016} have demonstrated that the excitonic temperature can deviate from the lattice temperature$-$often being higher due to non-equilibrium effects$-$in certain materials, no such measurements are currently available for MoS$_2$ bilayer. As a result, we use the lattice temperature as a practical approximation for the excitonic temperature. We note that since our calculations for PL are performed at room temperature (300 K), the actual excitonic temperature could be slightly elevated compared to the lattice temperature. However, we expect this difference to have only a minor impact on our results, especially given the relatively high thermal energy at this temperature.

\subsection{Supercell and finite-displacement approaches}
\label{sec:supercell}

We note that Eqs. \eqref{eq:rsp-} and \eqref{eq:rsp+} involve a sum over the phonon momentum $\mathbf{q}$. The unstrained bilayer MoS$_2$ exhibits an indirect band gap between $\Gamma$ (valence band maximum, VBM) and $\Lambda$ (conduction band minimum, CBM), as Sec. \ref{sec:mos2} shows below. Therefore, the leading phonon contributions will likely come from the phonons with momentum $\mathbf{q}=\Lambda-\Gamma$. In this work, we adopt a supercell approach proposed in Refs. \cite{lloyd2015,Paleari2019} and construct the smallest possible supercell such that the $\mathbf{q}$-phonon of the unit cell folds to the $\mq_{\rm sc}=\Gamma$ in the supercell (in this work, we add a subscript ``sc'' to denote quantities defined in the supercell). In this way, the indirect transition in the unit cell with a crystal momentum $\mathbf{q}$ becomes a direct $\mq_{\rm sc}=0$ transition in the supercell. We emphasize that this transformation does not change the oscillator strength: this transition remains optically dark, regardless of the simulation cell in which it is computed. The advantage of the supercell approach is two-fold: (i) one can describe the finite $\mathbf{q}$ valence-to-conduction transition with a standard BSE approach that does not involve finite exciton momenta; and (ii) one can describe the effects of the phonons using a finite-displacement approach where the atomic positions are explicitly distorted based on the phonon eigenvectors, because the $\mathbf{q}$-phonons of the unit cell are commensurate with the supercell.

For the unstrained bilayer MoS$_2$, $\Lambda=(1/6,1/6,0)$, expressed in fractional coordinates of the reciprocal lattice vectors. Therefore, we construct the following non-diagonal supercell:
\begin{equation}
\begin{pmatrix}
    \mathbf{a}_{1,\rm{sc}} \\ \mathbf{a}_{2,\rm{sc}}
\end{pmatrix}
=
\begin{pmatrix}
1 & 5 \\
0 & 6 \\
\end{pmatrix}
\begin{pmatrix}
\mathbf{a}_{1,\rm{uc}} \\ \mathbf{a}_{2,\rm{uc}}
\end{pmatrix}
\hspace{0.3in}
\mbox{for } \Lambda = (1/6,1/6,0).
\label{eq:ndsc1}
\end{equation}
Here, the subscript ``sc'' (``uc'') denotes quantities defined in the supercell (unit cell), and $\mathbf{a}_1$ and $\mathbf{a}_2$ are the first and second lattice vectors, respectively. $\mathbf{a}_3$ stays unchanged between the unit cell and the supercell.

The supercell defined in Eq. \eqref{eq:ndsc1} also applies to the case with 0.2\% and 0.6\% tensile strain, because under these strain conditions, the indirect gap remains between $\Gamma$ (VBM) and $\Lambda$ (CBM), qualitatively the same as the unstrained case (see Sec. \ref{sec:strain} below). At 1.4\% tensile strain, the conduction band at $K$ and $\Lambda$ are almost degenerate (see Sec. \ref{sec:strain} below). $K=(1/3,1/3,0)$ expressed in fractional coordinates of the reciprocal lattice vectors. Therefore, in addition to the supercell defined in Eq. \eqref{eq:ndsc1}, we consider another non-diagonal supercell:
\begin{equation}
\begin{pmatrix}
    \mathbf{a}_{1,\rm{sc}} \\ \mathbf{a}_{2,\rm{sc}}
\end{pmatrix}
=
\begin{pmatrix}
1 & 2 \\
0 & 3 \\
\end{pmatrix}
\begin{pmatrix}
\mathbf{a}_{1,\rm{uc}} \\ \mathbf{a}_{2,\rm{uc}}
\end{pmatrix}
\hspace{0.3in}
\mbox{for } K = (1/3,1/3,0).
\label{eq:ndsc2}
\end{equation}
For the 1.4\% tensile strain, the non-diagonal supercell defined in Eq. \eqref{eq:ndsc1} is used to compute and capture the $\mathbf{q}_1=\Lambda-\Gamma$ phonon contributions to PL, and the non-diagonal supercell defined in Eq. \eqref{eq:ndsc2} is used to compute and capture the $\mathbf{q}_2=K-\Gamma$ phonon contributions to PL, because the $K-\Gamma$ phonons of the unit cell fold to $\mq_{\rm sc}=0$ in this supercell. 

\begin{figure}[!ht]
\centering
\includegraphics[width=\linewidth]{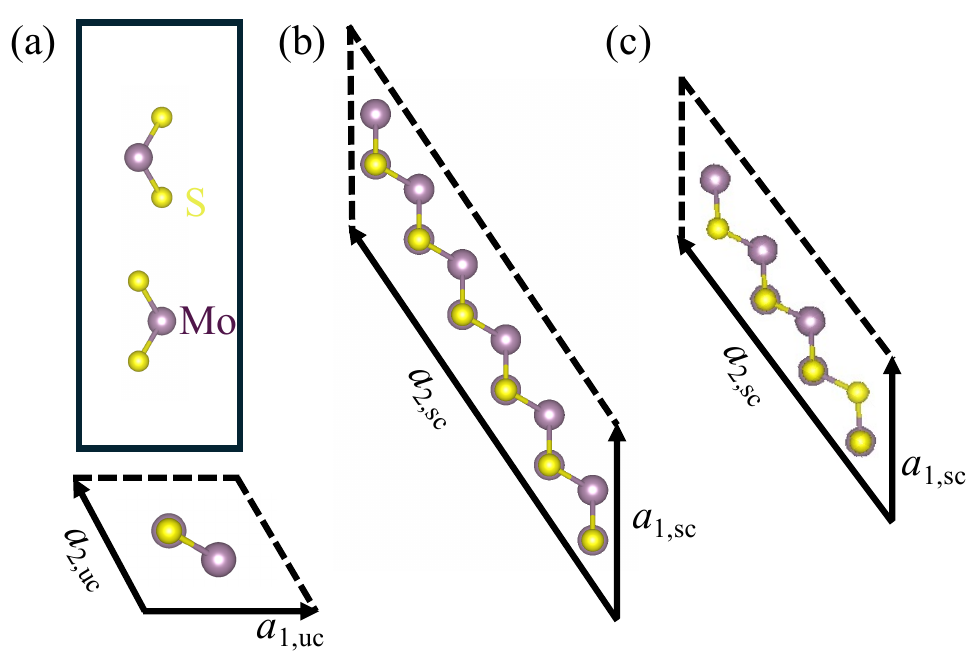}
\caption{Crystal structures of bilayer MoS$_2$ in the AA$^\prime$ stacking: (a) the unit cell in side (upper panel) and top (lower panel) view, (b) a non-diagonal supercell defined in Eq. \eqref{eq:ndsc1}, and (c) a non-diagonal supercell defined in Eq. \eqref{eq:ndsc2}.}
\label{fig:stru}
\end{figure}

Practically, for unstrained bilayer MoS$_2$ and 0.2\% and 0.6\% tensile strain, Eqs. \eqref{eq:rsp-} and \eqref{eq:rsp+} include one $\mq$ point, $\Lambda-\Gamma$ in the unit cell, and all the 18 phonon modes associated with it. Eq. \eqref{eq:finite-diff} requires BSE calculations of an undistorted supercell [defined using Eq. \eqref{eq:ndsc1}] and 18 distorted supercell structures. For 1.4\% tensile strain, Eqs. \eqref{eq:rsp-} and \eqref{eq:rsp+} include two $\mq$ points, $\Lambda-\Gamma$ and $K-\Gamma$, in the unit cell and the 18 phonon modes at each $\mq$. Eq. \eqref{eq:finite-diff} requires BSE calculations of two undistorted supercells [defined using Eqs. \eqref{eq:ndsc1} and \eqref{eq:ndsc2}] and 18 distorted structures in each supercell. The unit cell and the two supercells are schematically shown in Fig. \ref{fig:stru}.

In computing Eq. \eqref{eq:finite-diff}, $T(R_{\lambda\mq}=0)$ is computed in the supercell without distortions of atomic coordinates. For $T(R_{\lambda\mq})$, we distort the atomic coordinates in the supercell based on phonon eigenvectors, i.e., for the $\alpha$-th atom in the $l$-th unit cell within the supercell, its coordinates are displaced according to
\begin{equation}
R_{\alpha l,\lambda\mq} = A ~\mbox{Re}\left[\bm{e}_{\alpha,\lambda\mq}\exp\left(i\mq\cdot R_l\right)\right].
\label{eq:distort}
\end{equation}
Here, $\bm{e}_{\alpha,\lambda\mq}$ is the complex phonon eigenvector of mode $\lambda$ at wave vector $\mq$ (defined in the unit cell) for atom $\alpha$, and $R_l$ is the real-space lattice vector representing the position of the $l$-th unit cell in the supercell. The phase factor arises from Bloch's theorem for the phonons, reflecting how the atomic displacement pattern varies across the crystal. $A$ is the displacement amplitude and is chosen to be 0.01 Bohr in our calculations.

\subsection{Computational details}
\label{sec:comppara}

Density functional theory (DFT)-based calculations are performed using the \textsc{Quantum Espresso} \cite{QE} package. We employ the Perdew-Burke-Ernzerhof (PBE) functional \cite{PBE}, norm-conserving pseudopotentials from the \textsc{PseudoDojo} library \cite{Dojo2018}, a plane-wave kinetic-energy cutoff of 100 Ry, and a $\Gamma$-centered Monkhorst-Pack  \cite{Monkhorst1976} $\mathbf{k}$-mesh of $18\times18\times1$ for the bilayer MoS$_2$ unit cell. We keep a vacuum of about 21 \AA~along the $z$-direction. We include spin-orbit coupling (SOC) in all electronic structure calculations. The geometry relaxation is carried out by fixing the lattice parameter to be the experimental $a = 3.16$ \AA~\cite{Boker2001} and the Mo-Mo distance along the $z$ direction to be 6.41 \AA~\cite{Wilson1969}. We chose the latter so that the PBE indirect band gap is qualitatively correct (i.e., $\Gamma$ at VBM and $\Lambda$ at CBM). The dynamical matrix is computed using DFPT \cite{Baroni2001} on a $6\times 6 \times 1$ $\mathbf{q}$-mesh, without inclusion of SOC to reduce the computational cost.

The quasiparticle and optical properties are computed using $GW$-BSE as implemented in the \textsc{BerkeleyGW} package \cite{BGW}. Our calculations are $G_0W_0$@PBE, including 200 bands in the evaluation of the non-interacting Kohn-Sham polarizability $\chi^0$, along with a dielectric energy cutoff of 8.0 Ry and a $\mathbf{q}$-mesh of $18\times 18 \times 1$ for the bilayer MoS$_2$ unit cell. The $\mathbf{q}\rightarrow 0$ limit is addressed by including 67 bands on a shifted $\mathbf{q}_0$-point (0.00100, $-$0.00067, 0.00000) (expressed in fractional coordinates in the reciprocal space) in the calculation of $\chi^0$. The semiconductor screening and the slab Coulomb truncation schemes \cite{Ismail2006} are used. The Hybertsen-Louie generalized plasmon-pole model \cite{HL86} is used for the frequency dependence of the dielectric function. The static remainder approximation \cite{DSJC13} is used to accelerate the convergence of the quasiparticle energies.

%(for unitcell) For the BSE part, we include 10 valence and 10 conduction bands in computing the electron-hole interaction kernel, on a coarse $18\times 18 \times 1$ $\mathbf{k}$-mesh. The BSE Hamiltonian is constructed in an active space of 6 valence and 4 conduction bands, interpolated to a fine $36\times 36 \times 1$ $\mathbf{k}$-mesh before diagonalization. Finally, the optical absorption spectra are generated using an energy broadening of 25 meV.

The supercell defined in Eq. \eqref{eq:ndsc1} [Fig. \ref{fig:stru}(b)] contains six replicas of the unit cell. Consequently, we scale the $GW$-BSE parameters as follows: we include 1200 bands in the sum over states to compute $\chi^0$, together with a dielectric energy cutoff of 8.0 Ry and a $\mathbf{q}$-mesh of $18\times 3 \times 1$. For the $\mathbf{q}\rightarrow 0$ limit, we include 400 bands on a shifted $\mathbf{q}_0$-point (0.001, 0.001, 0.0), expressed in fractional coordinates in the reciprocal space. To reduce computational costs, we only calculate the $\chi^0$ explicitly for the undistorted supercell and use it repeatedly for all 18 phonon-distorted supercells. The effects of using $\chi^0$ of a perfect structure in materials with small deviations have been previously shown to be minimal \cite{Paleari2019, Li2023}. We include 26 valence and 26 conduction bands in the Coulomb interaction kernel, and construct the BSE Hamiltonian in an active space of 20 valence and 14 conduction bands, interpolated to a fine $36 \times 6 \times 1$ $\mathbf{k}$-mesh before diagonalization. We found that these parameters are sufficient to converge the results, i.e., achieving identical optical absorption spectra between the unit cell and the undistorted supercell up to 2.5 eV. The supercell defined in Eq. \eqref{eq:ndsc2} contains three replicas of the unit cell, and the computational parameters are scaled accordingly in a similar manner. The PL spectra computed using Eqs. \eqref{eq:rsp-} and \eqref{eq:rsp+} are generated using an energy broadening of 15 meV, unless noted otherwise.

\section{Results}

\subsection{Unstrained bilayer MoS$_2$}
\label{sec:mos2}

We consider the AA$^\prime$ stacking because it is the most stable configuration in our calculations and in the literature \cite{He2014, Mansouri2023}. As we mentioned in Sec. \ref{sec:comppara}, we fix the Mo-Mo distance along the $z$ direction to be 6.41 \AA~and use the experimental lattice parameter $a=3.16$ \AA~\cite{Boker2001} in our geometry relaxation of the atomic positions using the PBE functional. In Fig. S1 of the Supplemental Material \cite{SM}, we compare the PBE band structures with those from fully relaxed atomic structures (with the lattice parameter fixed as the experimental value) using the Grimme D2 van der Waals correction scheme \cite{Grimme-D2, Grimme-D2-2}. We can see that our constraint of the Mo-Mo distance along the $z$ direction leads to qualitatively correct results in the band gap and shows good agreement with prior studies \cite{Liu2014, Zhang2019}. 

\begin{figure}[!ht]
\centering	
\includegraphics[width=\linewidth]{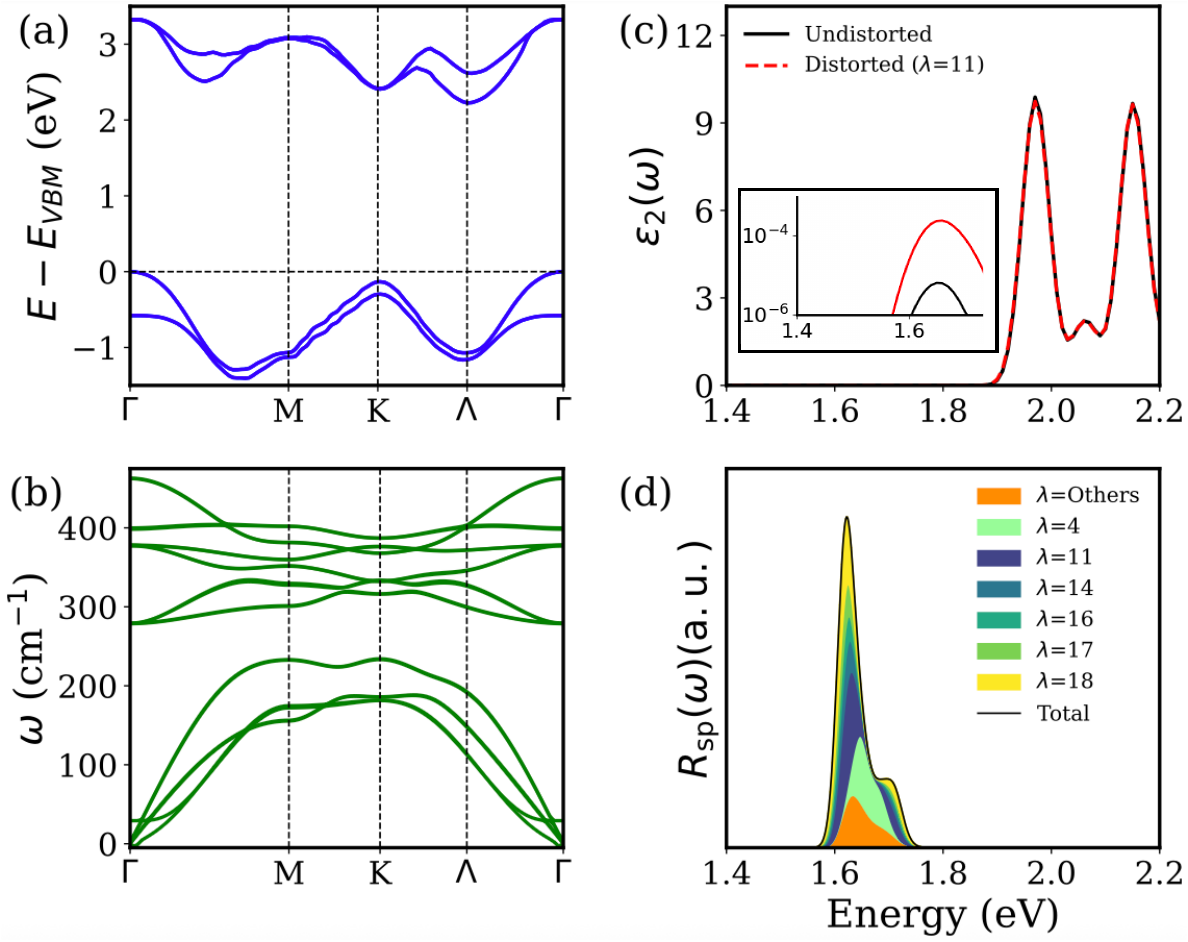}
\caption{(a) $G_0W_0$@PBE electronic band structure and (b) PBE phonon band structure of the AA$^\prime$-stacked bilayer MoS$_2$ unit cell. (c) BSE optical absorption spectra of the undistorted (black) and distorted (red) supercell of MoS$_2$ bilayer [Fig. \ref{fig:stru}(b)]. The atomic positions are distorted according to Eq. \eqref{eq:distort} for $\lambda=11$. The inset highlights the spectra plotted on a logarithmic scale to emphasize low-intensity features that give rise to the PL spectrum. (d) Calculated PL spectrum due to the indirect transition $\Lambda-\Gamma$, color-coded by the contributions from each phonon mode.}
\label{fig:mos2} 
\end{figure}

At the PBE level [Fig. S1(a)], the $\Gamma$ is the VBM, and the $\Lambda=(1/6,1/6,0)$ (expressed in fractional coordinates in the reciprocal space) is the CBM, with an indirect band gap of 1.45 eV. However, the $K$ point is nearly degenerate with $\Lambda$ at the CBM, with an energy of only 16 meV higher. Fig. \ref{fig:mos2}(a) presents the $G_0W_0$-corrected electronic band structure. At the $GW$ level, the $\Lambda-\Gamma$ indirect gap becomes 2.23 eV. The $G_0W_0$ further lifts the near degeneracy between the $K$ and $\Lambda$ points at the CBM, resulting in an energy difference of 180 meV between them and highlighting the significant impact of many-body corrections. These results agree with those of Ref. \cite{Hippalgaonkar2017}. Our $GW$ indirect ($\Lambda-\Gamma$) and direct ($K-K$) band gaps are converged within 0.02 eV and 0.04 eV, respectively. 

Fig. \ref{fig:mos2}(b) shows the calculated phonon dispersion of the bilayer MoS$_2$ unit cell, in very good agreement with prior studies \cite{Din2021, Bhatnagar2022}. Because the electronic band gap is indirect between $\Gamma$ and $\Lambda$, the leading phonon contribution to the PL is from those at the wave vector $\mq = \Lambda$ in the phonon dispersion. As shown in Fig. \ref{fig:mos2}(b), the phonon band structure exhibits no imaginary frequencies in the whole Brillouin zone, revealing its dynamical stability. The phonon bands can be categorized into two regions separated by an energy gap at about 250 cm$^{-1}$. The modes between 280 cm$^{-1}$ and 310 cm$^{-1}$ consist of contributions from S atoms only. Above (below) this range, we see larger contributions to the phonon modes from the vibrations of S (Mo) atoms, as shown in Fig. S2(a) in the Supplemental Material \cite{SM}.

As we discussed in \ref{sec:supercell}, we construct a non-diagonal supercell as defined in Eq. \eqref{eq:ndsc1} to compute the phonon-assisted PL spectra based on finite displacement of the phonons. For the bilayer MoS$_2$ unit cell, there are 6 atoms with 18 phonon modes. For the supercell defined in Eq. \eqref{eq:ndsc1} [shown in Fig. \ref{fig:stru}(b)], we therefore consider 18 different distortion patterns determined by Eq. \eqref{eq:distort}, labeled by the phonon mode $\lambda$ at wave vector $\mq=\Lambda-\Gamma$. A finite-difference evaluation of Eq. \eqref{eq:finite-diff} then requires $GW$-BSE calculations of 19 systems, one undistorted and 18 distorted. To reduce the computational cost, we introduce two approximations. First, we only perform an explicit $GW$ calculation for the undistorted supercell, and then fit the quasiparticle energies of the top 26 valence bands and bottom 26 conduction bands, respectively, to their PBE eigenvalues \cite{HL86}:
\begin{subequations}
\begin{align}
\epsilon^{\rm QP}_i & = c_1 \epsilon^{\rm PBE}_i + c_2 \\
\epsilon^{\rm QP}_a & = c_3 \epsilon^{\rm PBE}_a + c_4  
\end{align}
\label{eq:fit}
\end{subequations}
In these equations, $\epsilon^{\rm QP}$ is a quasiparticle energy, $\epsilon^{\rm PBE}$ is a PBE eigenvalue, $i$ is a valence band, and $a$ is a conduction band. $c_1$ to $c_4$ are fitting parameters from the method of least squares. Although we did not explicitly write out the $\mk$ index, the fitting of Eq. \eqref{eq:fit} is carried out for each $\mk$ point separately. Then we apply the fitting equations to each of the 18 distorted supercell structures to obtain $\epsilon^{\rm QP}$ from corresponding PBE calculations. 

As a second approximation, in the BSE calculations of the 18 distorted supercell structures, we use the same $\chi^0$ of the undistorted supercell to compute the electron-hole interaction kernel, as we mentioned in Sec. \ref{sec:comppara}. Note that the Kohn-Sham orbitals of the 18 distorted structures are different and calculated explicitly. To validate these two approximations, we perform benchmark calculations of the undistorted and selected distorted supercell structures by comparing them to direct BSE calculations. The excellent agreement in the optical absorption spectra shown in Fig. S3 of the Supplemental Material \cite{SM} confirms the reliability of our approach. 

Fig. \ref{fig:mos2}(c) shows the BSE optical absorption spectra of the undistorted supercell (black) and that of the supercell distorted along phonon mode $\lambda=11$ (red), which is selected due to its dominant contribution to the phonon-assisted PL (see below). The absorption spectrum of the undistorted supercell features the first main peak at 1.97 eV, with the second peak about 0.16 eV higher, in consistency with both experimental \cite{Mak2010} and theoretical \cite{Molina2013} results. These two peaks originate from the optical transitions from the VBM and VBM-1, respectively, to the CBM at the $K$ point of the Brillouin zone [c.f. Fig. \ref{fig:mos2}(a)]. The $GW$ $K-K$ band gap is 2.54 eV, indicating that the exciton binding energy is 0.57 eV, in good agreement with Refs. \cite{Cheiwchanchamnangij2012,Gerber2019}. With finite displacement of atomic positions due to phonons, the main peaks in the absorption spectrum (bright excitations) are largely unchanged. However, the dark excitations are significantly affected. This is the case for each phonon mode.

The inset in Fig. \ref{fig:mos2}(c) displays the absorption peaks on a logarithmic scale to highlight the low-intensity features of the dark excitations: for the undistorted supercell (black), this peak at about 1.65 eV corresponds to the $\mq=\Lambda-\Gamma$ indirect transition in the unit cell ($\mq_{\rm sc}=\Gamma$ in the supercell), which is dark due to momentum conservation. With explicit atomic distortions in the supercell along phonon eigenvectors as in the finite displacement approach, this $\mq_{\rm sc}=\Gamma$ transition becomes less dark. In fact, the transition matrix element is several orders of magnitude larger, as shown by the red dashed line. Physically, this is a manifestation of exciton-phonon coupling, described in terms of the change in the transition matrix element at the $\mq_{\rm sc}=\Gamma$ point in the supercell. Via Eqs. \eqref{eq:finite-diff}, \eqref{eq:rsp-}, and \eqref{eq:rsp+}, this change in the transition matrix element gives rise to PL intensity, which we plot in Fig. \ref{fig:mos2}(d). 
 
Fig. \ref{fig:mos2}(d) presents the total PL spectrum calculated using the sum of Eqs. \eqref{eq:rsp-} and \eqref{eq:rsp+}. It features two peaks: the one lower in energy arises from a phonon emission process [Eq. \eqref{eq:rsp+}], where the peak position ($E^S-\Omega_{\lambda\mq}$) is red-shifted compared to the phonon-less transition; the one higher in energy arises from phonon absorption [Eq. \eqref{eq:rsp-}], where the peak position ($E^S+\Omega_{\lambda\mq}$) is blue-shifted compared to the phonon-less transition. The energy separation between these two peaks corresponds to approximately twice the maximum phonon energy involved \cite{Bebb1972}. The inclusion of both phonon emission and phonon absorption is important in computing PL \cite{Cannuccia2019, Libbi2022}, especially at high temperatures \cite{Mak2010}, as we elaborate below in Sec. \ref{sec:temp-dep}.

\begin{figure}[!ht]
\centering	
\includegraphics[width=\linewidth]{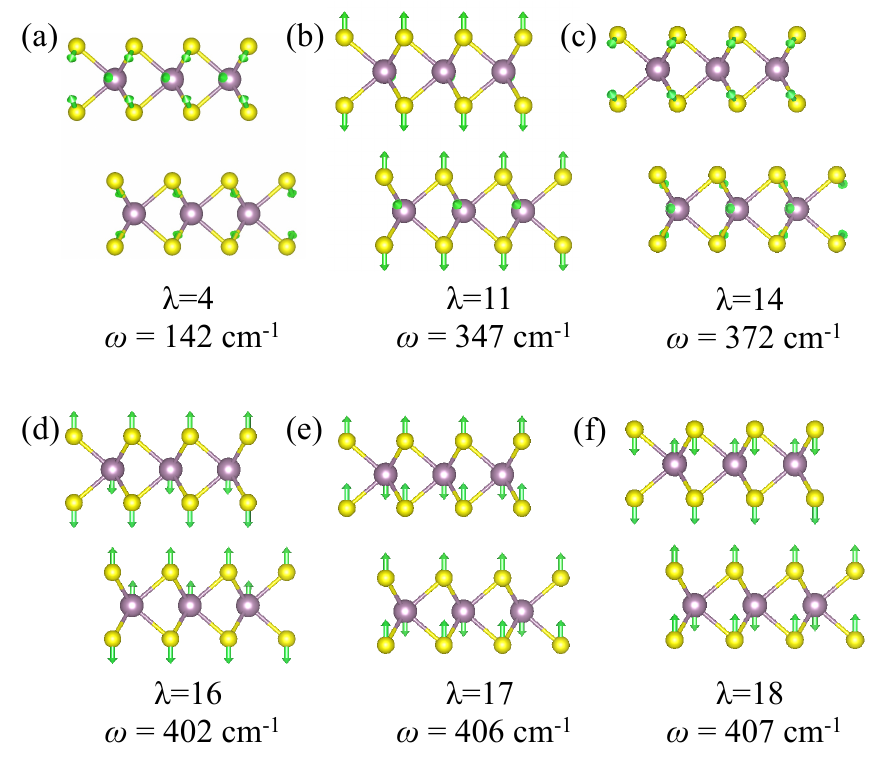}
\caption{Atomic displacement patterns of leading phonon modes that contribute strongly to the PL intensity: $\lambda=$ (a) 4, (b) 11, (c) 14, (d) 16, (e) 17, and (f) 18 at $\mathbf{q}=\Lambda-\Gamma$. The corresponding phonon frequencies $\omega$ are also indicated. The purple (yellow) color represents the Mo (S) atom.}
\label{fig3:disp} 
\end{figure}

We color-code Fig. \ref{fig:mos2}(d) by the contributions of leading optical phonon modes. This is possible because Eqs. \eqref{eq:rsp-} and \eqref{eq:rsp+} involve a sum over phonon modes. The mode $\lambda=11$, characterized by out-of-plane vibrations of S atoms combined with in-plane vibrations of Mo atoms [Fig. \ref{fig3:disp}(b)], contributes to about 23\% of the PL intensity. The percentage contribution is quantified by evaluating the area under its curve relative to the total area. The other leading contributions are listed in Table \ref{table2} and the atomic displacement patterns for these modes are shown in Fig. \ref{fig3:disp}. It is worth mentioning that the phonon modes between 280 and 310 cm$^{-1}$ -- which are associated solely with the vibrations of S atoms -- contribute minimally to the PL intensity. This arises because the band-edge states in bilayer MoS$_2$ are dominated by Mo $d$ orbitals \cite{Du2018,Padilha2014,Sun2019}, hence, S-only phonons do not couple strongly to these states. We also note that acoustic phonon modes contribute negligibly to the PL intensity. 

Our results are in excellent agreement with experimental measurements \cite{Mak2010, Conley2013, Zhu2013}, which reported a PL peak near 1.6 eV that is assigned to indirect transitions. Our computed width, however, is much narrower than the experiment. Part of the reason is that we have used a small energy broadening (15 meV) to broaden the delta functions in Eqs. \eqref{eq:rsp-} and \eqref{eq:rsp+}. More importantly, in our calculations, Eqs. \eqref{eq:rsp-} and \eqref{eq:rsp+} only involve one $\mq$ point to make the calculations tractable, while in reality, indirect transitions involve all possible $\mq$ points, leading to many phonon-modulated transitions $E^S\pm\Omega_{\lambda\mq}$. Additional experimental factors, such as substrates \cite{Buscema2014, Farhat2022, Ling2014, Yan2015, Tebyetekerwa2020}, might lead to additional mechanisms for broadening. 

Lastly, we note that by construction [we only considered the second term in Eq. \eqref{eq:taylor}], our approach does not capture the PL intensity from the direct and phonon-less emission, which is identified to be at about 1.85 eV experimentally \cite{Mak2010, Conley2013, Zhu2013}. However, we can perform a rough estimate: the calculated direct $K-K$ gap is 2.54 eV and the calculated indirect $\Lambda-\Gamma$ gap is 2.23 eV in $GW$. If we consider the exciton-phonon coupling is similar in magnitude between these two types of transitions, adding this difference in band gaps (0.31 eV) to our PL peak position (arising from $\Lambda-\Gamma$ transition, 1.6-1.7 eV) leads to an estimate of the direct PL peak to be 1.9-2.0 eV. 

\subsection{Temperature dependence}
\label{sec:temp-dep}

\begin{figure}[!ht]
\centering	
\includegraphics[width=\linewidth]{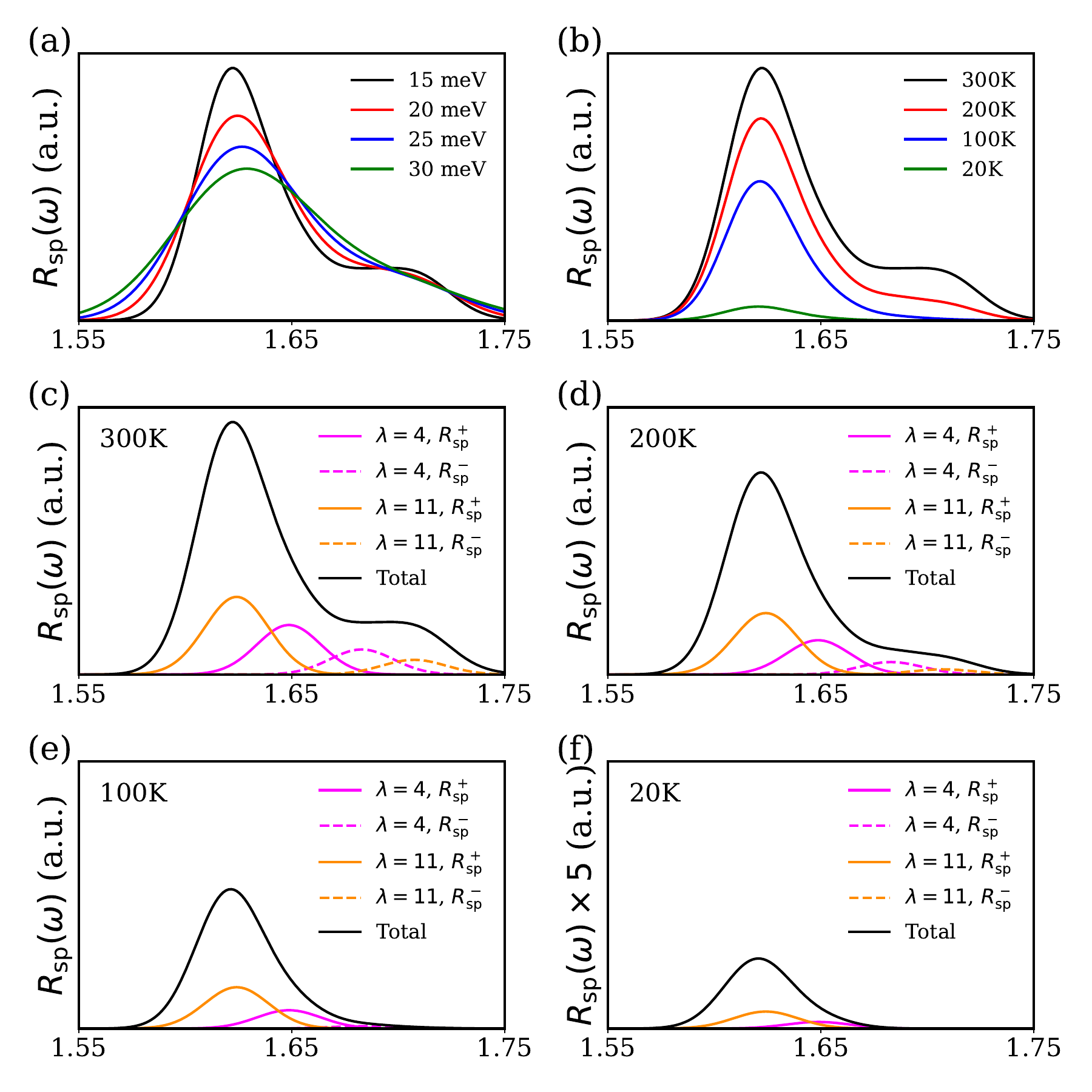}
\caption{(a) Fig. \ref{fig:mos2}(d) replotted with different energy broadening to approximate the delta functions in Eqs. \eqref{eq:rsp-} and \eqref{eq:rsp+}. (b) Computed PL spectra of bilayer MoS$_2$ at different temperatures, using an energy broadening of 15 meV (same below). Phonon mode $\lambda=4$ (magenta) and $\lambda=11$ (orange) contributions to $R_{\rm sp}^+(\omega)$ (solid) and $R_{\rm sp}^-(\omega)$ (dashed), as well as the total $R_{\rm sp}(\omega)$ from all phonon modes (solid black) at (c) 300 K, (d) 200 K, (e) 100 K, and (f) 20 K. In panel (f), the y-axis is scaled by a factor of 5 for clarity.}
\label{fig4:temp} 
\end{figure}

Before we examine the explicit temperature dependence that enters into Eqs. \eqref{eq:rsp-} and \eqref{eq:rsp+} through $n_B$ and $N_{\rm xct}$, we discuss the effect of energy broadening that approximates the delta functions in Eqs. \eqref{eq:rsp-} and \eqref{eq:rsp+}. Fig. \ref{fig4:temp}(a) re-plots the black curve in Fig. \ref{fig:mos2}(d), the total PL spectrum $R_{\rm sp}(\omega)$, but with different values for the energy broadening, at $T=300$ K. With an energy broadening of 15 meV, the PL spectrum consists of two peaks, one arising from phonon emission and one arising from phonon absorption. The latter is higher in energy but lower in intensity: the PL with phonon absorption [Eq. \eqref{eq:rsp-}] is proportional to $\delta(\omega-E^S-\Omega_{\lambda\mq})n_B$ while the PL with phonon emission [Eq. \eqref{eq:rsp+}] is proportional to $\delta(\omega-E^S+\Omega_{\lambda\mq})(n_B+1)$. Therefore, the intensity of the peak due to phonon emission is always higher. When the value of the energy broadening increases, the widths of both peaks increase, and the two peaks merge into a broader peak with a larger width. Physically, a larger width may be associated with larger structural variations resulting from a higher temperature. 

Fig. \ref{fig4:temp}(b) shows the total PL spectrum $R_{\rm sp}(\omega)$ at different temperatures: 300 K, 200 K, 100 K, and 20 K. At a lower temperature, both peaks are smaller, due to a decreased phonon population $n_B$ for each phonon mode. At low enough temperature, $n_B$ is nearly zero for all phonon modes, and only the peak due to phonon emission (proportional to $n_B+1$) is visible, as the green curve in Fig. \ref{fig4:temp}(b) shows. As a result, the absence of the phonon absorption process in PL at low temperatures leads to a sharper PL spectrum. This behavior is consistent with earlier studies \cite{Du2018, Pei2015}, which demonstrated that reducing the temperature leads to a narrower and sharper PL peak. We note that Ref. \cite{Paleari2019} only focused on the phonon-emission contribution to PL, $R_{\rm sp}^+(\omega)$, due to the low temperature considered (10 K) for hBN. Here, by considering a broad range of temperatures, we elucidate how both phonon-absorption and phonon-emission processes evolve with temperature. 

The solid black lines in Fig. \ref{fig4:temp}(c)-(f) re-plot the four curves in Fig. \ref{fig4:temp}(b), with one temperature value in each panel. We further consider two representative phonon modes ($\lambda=4$ and $\lambda=11$, see Fig. \ref{fig3:disp}), and discuss the difference in their temperature-dependent behaviors. We choose these two modes due to their large contributions to PL [see Table \ref{table2} and Fig. \ref{fig:mos2}(d) and discussions therein], and more importantly, their different frequencies. We plot their contributions to $R_{\rm sp}^+(\omega)$ and $R_{\rm sp}^-(\omega)$ separately to clearly show the peaks from both phonon-absorption and phonon-emission processes. We can see that for both modes, the peak height becomes lower at a lower temperature. At room temperature, $dn_B/dT$ is larger for $\lambda=4$, such that its contribution to PL is more susceptible to the temperature change.

\subsection{Under tensile strain}
\label{sec:strain}

Experimentally, the electronic, vibrational, and optical properties of MoS$_2$ are known to be sensitive to external strain \cite{Buscema2014,Niehues2019}. As far as PL is concerned, strain can induce notable shifts in the emission energy and modify the relative intensities of indirect transitions \cite{Conley2013, Zhu2013,Qiao2022}. Recent studies reveal that biaxial strain and hydrostatic pressure can markedly reshape the excitonic landscape in MoS$_2$, where the strain activates otherwise forbidden phonon-assisted recombination channels to enhance PL efficiency \cite{Saraswat2025}.

Here, we apply uniaxial (100) tensile strain to bilayer MoS$_2$ at three different levels: 0.2\%, 0.6\%, and 1.4\%. Similar to the unstrained case, the atomic structures are relaxed using the PBE functional with the interlayer Mo-Mo distance fixed. As shown in Fig. S1 in the Supplemental Material \cite{SM}, the resulting band structure qualitatively agrees with that from a full geometry relaxation using Grimme D2 \cite{Grimme-D2, Grimme-D2-2} (with the lattice parameter fixed as the experimental value).

\begin{figure}[!ht]
\centering	
\includegraphics[width=\linewidth]{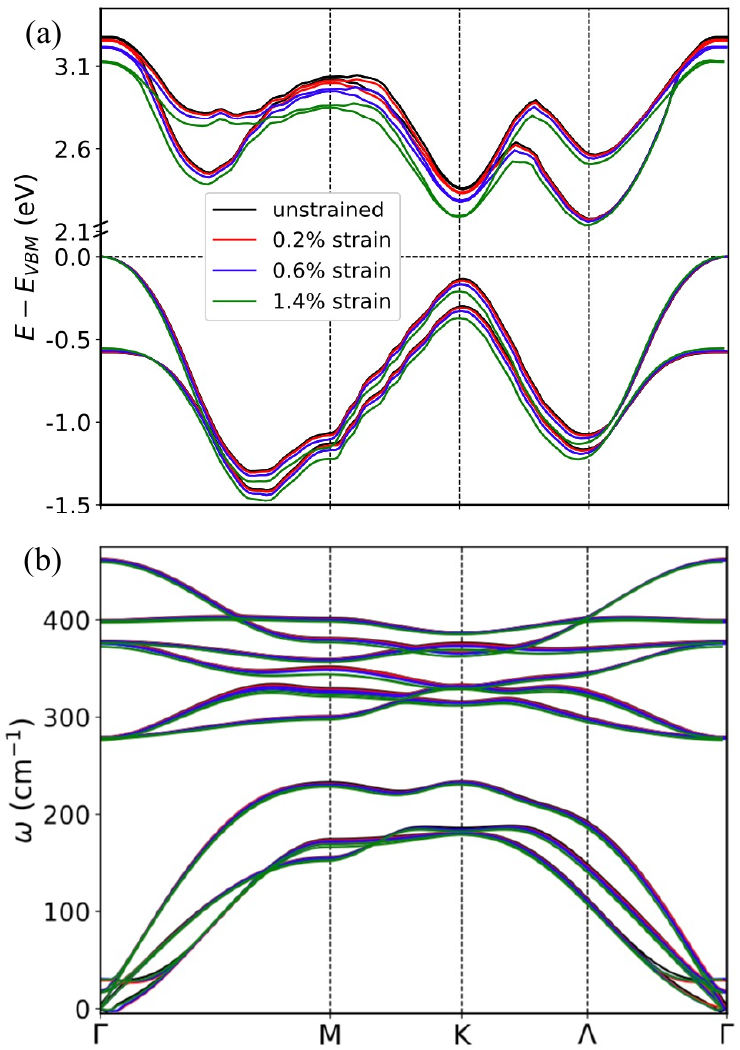}
\caption{(a) $G_0W_0$@PBE electronic band structure and (b) PBE phonon dispersion of bilayer MoS$_2$ with various uniaxial strain: unstrained (black), 0.2\% strain (red), 0.6\% strain (blue), and 1.4\% strain (green).}
\label{fig5:strain-prop} 
\end{figure}

Fig. \ref{fig5:strain-prop}(a) shows the $G_0W_0$@PBE electronic band structure with different strain, with key band gaps summarized in Table \ref{table1}. We see that the VBM energy at $\Gamma$ and the CBM energy at $\Lambda$ do not significantly change as strain, leading to an almost constant $E_g^{\Lambda-\Gamma}$ in Table \ref{table1}. However, the conduction band at $K$ decreases monotonically as the strain level increases, leading to a decreasing $E_g^{K-\Gamma}$. At 1.4\% strain, the $K$ and $\Lambda$ at CBM are almost degenerate, with only 50 meV energy difference (we emphasize that the CBM is still located at $\Lambda$). Therefore, at 0.2\% and 0.6\% strain, we only consider the $\mq=\Lambda-\Gamma$ phonons, same as the unstrained case. However, at 1.4\% strain, it is necessary to consider the phonon contributions at two $\mq$ points: $\Lambda-\Gamma$ and $K-\Gamma$. The former (latter) is commensurate with the supercell defined in Eq. \eqref{eq:ndsc1} [Eq. \eqref{eq:ndsc2}] and schematically shown in Fig. \ref{fig:stru}(b) [Fig. \ref{fig:stru}(c)]. Correspondingly, when we compute Eqs. \eqref{eq:rsp-} and \eqref{eq:rsp+} for the PL intensity, we consider two sets of phonon modes at two $\mq$ points for 1.4\% strain. Later, we will show that the $\mq=K-\Gamma$ phonons open up new channels for PL. We also note that the VB splitting at the $K$ point (due to a combination of SOC and interlayer interaction \cite{Cheiwchanchamnangij2012}) remains about 0.16 eV for all strain levels.

Fig. \ref{fig5:strain-prop}(b) shows the phonon band structure with different strain, with the phonon density of states shown in Fig. S2 in the Supplemental Material \cite{SM}. We can see that the effect of strain on the phonon band structure is smaller than that on the electronic band structure, and the S-only modes between 280 cm$^{-1}$ and 310 cm$^{-1}$ as found in the unstrained case remain present under strain. Quantitatively, our calculations reveal that under strain, the phonon modes undergo a moderate softening. For example, at 1.4\% strain, a few of the highest phonon modes exhibit a softening of about 5 cm$^{-1}$ ($\sim 0.6$ meV) at $\Lambda$ and about 8 cm$^{-1}$ ($\sim 1$ meV) at $K$, compared to the unstrained system. 

\begin{table}[h!]
\centering
\caption{Electronic and optical properties of unstrained and strained bilayer MoS$_2$. $E_g$ is the $G_0W_0$@PBE band gap, where the superscript shows the transition between the valence and conduction bands. $E_s$ is the position of the first absorption peak in $\varepsilon_2(\omega)$. $\bar{\omega}$ is the phonon-less indirect electronic transition that enters into the PL formalism, which equals the $E^S$ in the delta functions of Eqs. \eqref{eq:rsp-} and \eqref{eq:rsp+}. All values are in eV.}
\begin{tabular}{cccccc}
\hline
strain & $E_g^{\Lambda-\Gamma}$ & $E_g^{K-\Gamma}$ & $E_g^{K-K}$ & $E_s$ & $\bar{\omega}$ \\ 
\hline
     unstrained   & 2.23     & 2.41       & 2.54      & 1.97         & 1.66 \\
     0.2\%        & 2.22     & 2.38       & 2.52      & 1.95         & 1.65 \\
     0.6\%        & 2.21     & 2.33       & 2.50      & 1.93         & 1.64 \\
     1.4\%        & 2.19     & 2.24       & 2.45      & 1.89         & 1.71 \\
\hline
\end{tabular}
\label{table1}
\end{table}

\begin{figure}[!ht]
\centering
\includegraphics[width=\linewidth]{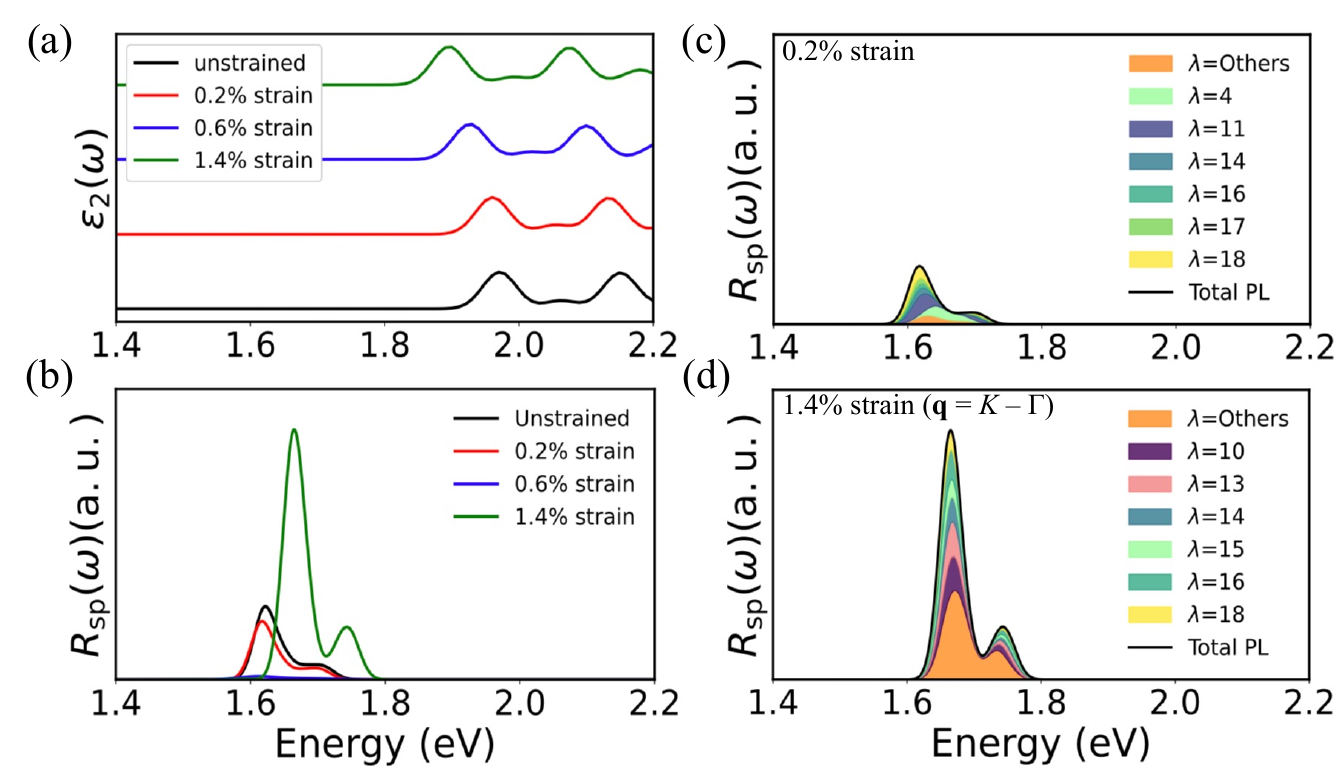}
\caption{(a) Optical absorption spectra from BSE and (b) total PL spectra for unstrained and strained bilayer MoS$_2$. (c) PL contributions from the indirect $\Lambda-\Gamma$ transition at 0.2\% strain. (d) PL contributions from the indirect $K-\Gamma$ transition at 1.4\% strain. Different color coding shows the contributions from various phonon modes to the PL peak.}
\label{fig6:strain-pl} 
\end{figure}

Fig. \ref{fig6:strain-pl}(a) compares the BSE optical absorption spectra for the unstrained and strained bilayer MoS$_2$. With the level of strain increasing, there is a noticeable red shift in the peak positions. This is well correlated with the $E_g^{K-K}$ in Table \ref{table1}, as these peaks arise from bright excitations at the $K$ point. In fact, the difference between $E_g^{K-K}$ and $E_s$ in Table \ref{table1} stays unchanged, suggesting that the exciton binding energy is unaffected by the strain.

Fig. \ref{fig6:strain-pl}(b) compares the total PL spectra for different levels of strain. The PL spectrum for 0.2\% strain resembles that of the unstrained material, with slightly lower intensity. This lowered intensity can be attributed to a decrease in $N_{\rm xct}$ in Eqs. \eqref{eq:rsp-} and \eqref{eq:rsp+}, due to the excited states being further apart [i.e., larger $E^S-E^{S1}$ in Eq. \eqref{eq:xct-occu}] under 0.2\% strain than in the unstrained case. Going from 0.2\% to 0.6\% strain, the PL intensity is greatly reduced [the blue line in Fig. \ref{fig6:strain-pl}(b) is almost invisible on this scale]. This can be attributed to a decrease in $\partial^2\left|T^S\right|^2/\partial R_{\lambda\mathbf{q}}^2$. This decrease in PL intensity is consistent with experimental observations in Refs. \cite{Conley2013, Zhu2013}. Furthermore, we observe that from the unstrained case to 0.6\% strain, the shape of the PL spectrum and the peak positions remain nearly unchanged. $\bar{\omega}$ in Table \ref{table1} represents the phonon-less indirect electronic transition energy. Recall $\bar{\omega}=(\omega_e+\omega_a)/2$ is numerically the $E^S$ in the delta functions in Eqs. \eqref{eq:rsp-} and \eqref{eq:rsp+} and also the midpoint between the two PL peaks. For these strain levels, this quantity stays unchanged, which can be attributed to the nearly unchanged $E_g^{\Lambda-\Gamma}$.

When the level of strain increases to 1.4\%, however, we observe a significant intensity increase and blueshift in peak positions. This complete change in profile is due to additional phonon channels becoming available at 1.4\% strain. Recall in Fig. \ref{fig5:strain-prop}(a), at 1.4\% strain, the $K$ and $\Lambda$ in the conduction band are nearly degenerate, leading us to consider two phonon momenta ($\mq=\Lambda-\Gamma$ and $\mq=K-\Gamma$) and correspondingly two supercells [Eqs. \eqref{eq:ndsc1} and \eqref{eq:ndsc2}, or Figs. \ref{fig:stru}(b) and (c)]. We can verify that the $\mq=\Lambda-\Gamma$ contributions to the total PL are even smaller in magnitude than the blue curve in Fig. \ref{fig6:strain-pl}(b), with the shape and peak positions very similar to those of the unstrained, 0.2\% strain, and 0.6\% strain. The green curve in Fig. \ref{fig6:strain-pl}(b) is almost entirely from the $\mq=K-\Gamma$ contributions to the PL, suggesting a much higher exciton-phonon coupling at this phonon wave vector. The $\mq=K-\Gamma$ phonon contributions are negligible in lower levels of strain due to the elevated energy at $K$ in the conduction band. The corresponding $\bar{\omega}$ value in Table \ref{table1} is different from other strain levels for the same reason. 

\begin{table*}[hbt]
\centering
\caption{Leading contributions to $R_{\rm sp}(\omega)$ from phonon modes, arranged in descending order based on percentage contribution, for unstrained and strained bilayer MoS$_2$. For 1.4\% strain, we list contributions from two different $\mq$ points. Atomic displacement patterns of these modes are shown in Fig. \ref{fig3:disp} for the unstrained case, and Fig. S4 in the Supplemental Material \cite{SM} for the strained cases.}
\begin{tabular}{ccccc}
\hline
unstrained & 0.2\% strain & 0.6\% strain & 1.4\% strain ($\mq=\Lambda-\Gamma$) & 1.4\% strain ($\mq=K-\Gamma$) \\ 
\hline
$\lambda = 11$ (23 \%) & $\lambda = 11$ (24 \%) & $\lambda = 11$ (49 \%) & $\lambda = 11$ (44 \%) & $\lambda = 10$ (13 \%) \\
$\lambda = 4$ (19 \%) & $\lambda = 4$ (17 \%) & $\lambda = 12$ (28 \%)& $\lambda = 12$ (28 \%) & $\lambda = 13$ (12 \%)\\
$\lambda = 18$ (16 \%)& $\lambda = 18$ (17 \%)& $\lambda = 14$ (9 \%)& $\lambda = 17$ (7 \%) & $\lambda = 16$ (11 \%)\\
$\lambda = 17$ (9 \%) & $\lambda = 17$ (9 \%) & $\lambda = 15$ (7 \%)& $\lambda = 15$ (5 \%) & $\lambda = 14$ (9 \%)\\
$\lambda = 14$ (8 \%) & $\lambda = 14$ (8 \%) & $\lambda = 18$ (6 \%)& $\lambda = 18$ (5 \%) & $\lambda = 18$ (7 \%)\\
$\lambda = 16$ (7 \%) & $\lambda = 16$ (7 \%) & $\lambda = 13$ (1 \%)& $\lambda = 14$ (4 \%) & $\lambda = 15$ (6 \%)\\
\hline
\end{tabular}
\label{table2}
\end{table*}

The qualitative change in the PL spectrum at 1.4\% strain also suggests different phonon mode contributions. Fig. \ref{fig6:strain-pl}(c) [(d)] shows the color-coded phonon contributions to the PL spectrum at 0.2\% strain (at 1.4\% strain, $\mq=K-\Gamma$ only, as the $\mq=\Lambda-\Gamma$ contributions are negligible). Table \ref{table2} lists the leading contributions to $R_{\rm sp}(\omega)$ from different phonon modes for the unstrained and strained bilayer MoS$_2$, in descending order based on their percentage contributions. The atomic displacement patterns of these modes are shown in Fig. \ref{fig3:disp} for the unstrained case and Fig. S4 in the Supplemental Material \cite{SM} for the strained cases.

For the unstrained material, 0.2\% strain, 0.6\% strain, and the $\mq=\Lambda-\Gamma$ contributions to the PL at 1.4\% strain, the dominant phonon contributions are from optical modes involving out-of-plane S vibrations and in-plane Mo vibrations, such as the $\lambda=11$ mode in these four cases. In contrast, for the $\mq=K-\Gamma$ contributions to the PL at 1.4\% strain, the dominant contributions are from phonons involving in-plane S vibrations and out-of-plane Mo vibrations, such as $\lambda=10$ and $\lambda=13$ (Fig. S4 in the Supplemental Material \cite{SM}). This distinction is a manifestation of the interplay between phonon momentum, vibrational patterns, and excitons. As a result, at 1.4\% strain, the indirect PL in bilayer MoS$_2$ originates from a different mechanism than in a lower level of strain.

\section{Conclusion}
\label{sec:conclude}

In this work, we employed first-principles $GW$-BSE to study the phonon-assisted PL of bilayer MoS$_2$, focusing on its temperature and strain dependence. Using the van Roosbroeck-Shockley relationship, the phonon-assisted PL intensity is expressed in terms of the changes in optical absorptions when the atoms in the system are explicitly displaced along phonon eigenmodes. The effects of exciton-phonon coupling are taken into account in a supercell approach, where we constructed supercells that are commensurate with the phonons of interest. For the unstrained material, 0.2\% strain, and 0.6\% strain, we only included $\mq=\Lambda-\Gamma$ phonons, because the VBM is located at the $\Gamma$ point and the CBM is located at the $\Lambda$ point in the $GW$ band structure. At 1.4\% strain, due to the near-degeneracy between the $K$ and $\Lambda$ points in the conduction band, we considered two $\mq$ points, $\mq=\Lambda-\Gamma$ and $\mq=K-\Gamma$. As a result, we constructed two types of supercells to include phonons at these two momenta. We included both phonon-emission and phonon-absorption processes in our calculations of the PL spectra, and showed that both contributions decrease when the temperature is lower, and only the PL peak due to phonon emission is visible at a low enough temperature. When $\mq=\Lambda-\Gamma$ is the only phonon momentum available, the PL intensity decreases as the strain level increases, with the peak positions and lineshape largely unchanged. However, at 1.4\% strain, the PL intensity increased and the peaks are blue-shifted, due to the availability of $\mq=K-\Gamma$ phonons at this strain level. Analysis of the mode-resolved contributions to PL showed that for $\mq=\Lambda-\Gamma$ phonons, the dominant contributions are from out-of-plane S vibrations and in-plane Mo vibrations, while for $\mq=K-\Gamma$ phonons, the dominant contributions are from in-plane S vibrations and out-of-plane Mo vibrations. Our findings highlight the rich interplay between strain, excitons, and phonons in bilayer MoS$_2$, and provide insights into the exciton-phonon coupling and phonon-mediated optical processes in similar layered materials.
\newline

\begin{acknowledgments}

We thank Fulvio Paleari for insightful discussions on the theory of PL, and thank Felipe da Jornada for the discussion of transition matrix elements in BSE. This work was supported by the U.S. Department of Energy (DOE), Office of Science, Basic Energy Sciences, under award no. DE-SC0023324, and used resources of the National Energy Research Scientific Computing Center (NERSC), a DOE Office of Science User Facility supported by the Office of Science of the U.S. DOE under Contract No. DE-AC02-05CH11231 using NERSC award BES-ERCAP0031564. Z.-F.L. also acknowledges an Alfred P. Sloan Research Fellowship (FG-2024-21750).

\end{acknowledgments}

\bibliography{ref.bib}
\end{document}